\documentclass[natbib]{svjour3}
\usepackage{natbib}
\usepackage{graphicx}

\begin{document}

\title{Magnetic fields in galactic haloes}

\author{M. Haverkorn$^1$ \and
        V. Heesen$^2$}

        \institute{M. Haverkorn \at ASTRON, PO Box 2 7990 AA Dwingeloo, the Netherlands; \at Leiden Observatory, Leiden University, PO Box 9513, 2300 RA Leiden, the Netherlands
        \and V. Heesen \at Centre for Astrophysics Research, University of Hertfordshire, Hatfield, AL 10 9AB, United Kingdom}
\date{Received: date / Accepted: date}

\maketitle

\begin{abstract}

Magnetic fields on a range of scales play a large role in the
ecosystems of galaxies, both in the galactic disk and in the extended
layers of gas away from the plane. Observing magnetic field strength,
structure and orientation is complex, and necessarily
indirect. Observational data of magnetic fields in the halo of the
Milky Way are scarce, and non-conclusive about the large-scale
structure of the field. In external galaxies, various large-scale
configurations of magnetic fields are measured, but many uncertainties
about exact configurations and their origin remain. There is a strong
interaction between magnetic fields and other components in the
interstellar medium such as ionized and neutral gas and cosmic
rays. The energy densities of these components are comparable on large
scales, indicating that magnetic fields are not passive tracers but
that magnetic field feedback on the other interstellar medium
components needs to be taken into account.

\end{abstract}

\section{Introduction}

Significant magnetic fields are present not only in galaxy disks, but
also in the gaseous galactic halo\footnote{The term 'galactic halo' is
used here to describe the medium above the galactic disk, also
sometimes referred to as the thick disk. In this definition, the
gaseous halo of a galaxy has not necessarily a direct relationship to
the spherical stellar halo.}. Magnetic fields in these galactic haloes
can provide a significant pressure component to counteract the gravity
of the halo gas \citep{bc90}. Also, they play a major role in the
disk-halo interaction, influencing superbubble break-out and funneling
particles out to intergalactic space. Finally, the determination of
the morphology of halo magnetic fields can shed light on the validity
of various dynamo models to explain the origin and evolution of
galactic magnetism.

We will first give a brief summary of the conditions in the haloes of
the Milky Way and nearby spirals in Section~\ref{s:gas}
and~\ref{s:cosmicray}. Sections~\ref{s:bfield} to~\ref{s:externals}
describe magnetic field strengths, scale heights, and structure in the
Milky Way and external spiral galaxies. Finally, in
Section~\ref{s:interaction} we briefly discuss the interaction of
magnetic fields with other interstellar medium components.

\section{Gaseous components in galactic haloes}
\label{s:gas}

\subsection{The Galaxy}

The interstellar medium (ISM) in the halo of the Milky Way consists of
the three-phase medium as in the disk, but with varying filling
factors and scale heights. While there is a warm neutral hydrogen
layer and cold neutral hydrogen clouds at large distances from the
Galactic plane \citep{dl90}, coupled to magnetic fields through charge
exchange, the most direct influence on magnetic field structure comes
from the halo plasma.

Estimates for the scale height of the Warm Ionized Medium (WIM) in the
Milky Way vary between $\sim$1000 and $\sim$1800~pc
\citep{r91,gmc08,sw09}. The volume filling factor of the WIM is
believed to increase from $f\sim0.1$ at the Galactic mid-plane to $f >
0.3$ at $|z|=$1000~pc \citep{bmm06}. Whereas H~{\sc II} regions
dominate the WIM distribution in the Galactic plane, at high latitudes
H$\alpha$ emission from the WIM is more diffuse and pervasive
\citep{hrt03}.

Hot, $\sim10^6$~K gas existing inside supernova remnants and
superbubbles is predicted to have a filling factor of about $\sim20\%$
\citep{f98}. This estimate is confirmed by the filling factor of large
neutral hydrogen holes \citep{h80} and observations of O~{\sc VI} UV
absorption lines \citep{sc94}. A patchy distribution of hot gas at
high latitudes is seen in fluctuations in the projected O~{\sc VI}
column densities \citep{hb96} and from X-ray shadows of interstellar
clouds.  Estimates of the hot gas scale height, from absorption lines
of high-ionization ions like O~{\sc VI} and N~{\sc V}, range from
about 3 to 5~kpc \citep{f01}.

\subsection{External galaxies}

Rossa \& Dettmar (2003a,b) reported a mean vertical extent of 1-2~kpc of the
diffuse ionized gas (DIG)\footnote{Following convention, we call the layer of
  ionized gas in spiral galaxies the Warm Ionized Medium (WIM) in the Milky
  Way, and Diffuse Ionized Gas (DIG) in external galaxies.} in external
spiral galaxies, if they possess warm dust as indicated by the FIR
$\lambda_{60}$/$\lambda_{100}$ ratio. This shows that gaseous haloes are
connected to the star formation activity in the underlying disk. External
galaxies also possess hot X-ray emitting gas with a scale height in the range
4-8~kpc \citep{strickland_04a}. \citet{tuellmann_00a} showed that this gas is
again correlated with the presence of warm dust, as for the extraplanar DIG.

\section{Cosmic rays}
\label{s:cosmicray}

\subsection{Scale height}

\paragraph{The Galaxy:} From estimates of the gamma-ray radiation from the
Compton Gamma-Ray Observatory in the $1-30$~MeV band, and isolating
the inverse Compton component, \citet{st96} concluded that the cosmic
ray scale height is probably much larger than that of the gas. An
alternative method to estimate the vertical scale height of cosmic
rays is based on counts of cosmic ray nuclei such as Li, Be, and B,
created in collisions of primary cosmic ray nuclei with interstellar
hydrogen. Using this method, \citet{bdd93} derived a cosmic ray scale
height $\le 3$~kpc at the solar radius, consistent with values found
by \citet{wlg92}. The cosmic ray equivalent scale height derived from
observed synchrotron emissivities in the Galaxy is about 2~kpc, in
agreement with the estimates above \citep{f01}.

Modeling of the Galactic synchrotron distribution from an all-sky
survey at 408~MHz \citep{hss82} shows that the synchrotron emissivity
in the Milky Way consists of a thick and a thin disk
\citep{bkb85}. The scale heights of these components are $\sim$150~pc
and $\sim$1500~pc, respectively (after rescaling to a Galactocentric
radius of 8.5~kpc), and 90\% of the total power is emitted in the
thick disk. The equipartition\footnote{The equipartition assumption
indicates equal energy densities in the magnetic field and cosmic
rays. It is often used to estimate magnetic field strengths from
measured synchrotron emissivities, and generally holds well on
galaxy-size scales.} value for the cosmic ray scale height derived
from this again roughly agrees with the above estimates.

\paragraph{External galaxies:} Similarly, in external galaxies vertical
profiles of the synchrotron emission can be characterized by two components, a
thin disk with a scale height of 300\,pc and a thick disk with a scale height
of 1.8\,kpc (at 4.8~GHz). These numbers are remarkably independent of the
star-formation rate \citep{krause_09a}. If we assume equipartition, the scale
height of the electrons would thus be 3.6~kpc. Not much is known if the
cosmic-ray nuclei and the electrons are transported together but this is
generally assumed. Therefore, for electrons with an energy of a few GeV, the
scale height in external galaxies agrees roughly with the ones quoted above
for the Galaxy.

\subsection{Cosmic-ray transport}

The luminosity of X-ray emission in the center of the Milky Way
can only be explained if the Galaxy possesses a wind that is hybridly
driven by cosmic rays and the thermal gas \citep{everett_08a}. Due to the
interaction with the ions and electron in the so-called streaming instability
the cosmic rays can push out the ionized gas by momentum transfer, and -- if
there are sufficient ion-neutrals collisions -- also the neutral gas. In a
study of the resolved emission in NGC\,253 it was shown that the scale height
of the electrons is proportional to their lifetime leading to an estimate of
300~$\rm km\,s^{-1}$ as the cosmic-ray bulk speed. This is similar to the
escape velocity in this galaxy and rather constant across the full extent of
the disk, so that a ``disk wind'' was proposed \citep{heesen_09a}.

\section{Magnetic field scale height and strength}
\label{s:bfield}

\subsection{Scale heights}

\paragraph{The Galaxy:} 
From hydrostatic equilibrium considerations, \citet{bc90} determined
that cosmic rays, magnetic fields, and a significant fraction of the
ISM should have scale heights in excess of 1~kpc. In particular, the
magnetic field strength only decreases by $\sim30\%$ out to a distance
of a kpc from the Galactic plane. A measure for the scale height of
the Galactic magnetic field can also be found from the synchrotron
scale height. \citet{bkb85} modeled synchrotron emissivity from
measurements of \citet{hks81, hss82}, which indicates a magnetic field
scale height at the location of the Sun of about 4.5~kpc
\citep{f01}. This scale height agrees with estimates considering
hydrostatic equilibrium in the halo \citep{kk98}. The discrepancy with
the smaller scale height of the regular magnetic field component found
from Faraday rotation measurements was explained by \citet{bc90} by
stating that the Galactic magnetic field becomes less regular away
from the plane. Alternatively, these measurements may be explained by
two separate magnetic field components with different scale heights
\citep{am88, hq94}, consistent with observations of a two-layer
structure of the Galactic non-thermal synchrotron emission
\citep{bkb85}.

\paragraph{External galaxies:}
The scale heights in external galaxies can be estimated from the
synchrotron scale height. If we assume equipartition, we obtain that
the magnetic field scale height is a factor of 4 larger than the
synchrotron scale height and thus about 6~kpc.

\subsection{Field strengths}

\paragraph{The Galaxy:} 
Much is still unknown about the magnetic field strength and structure
in the Milky Way at large distances from the Galactic plane.
\citet{srw08} modeled the magnetic field in the Galactic halo as a
plane parallel field superimposed on a turbulent component, based on
synchrotron emissivity and Faraday rotation data. They found that the
halo field is unrealistically high with a 1~kpc scale height, but when
using a scale height of 1.8~kpc, their estimate of the regular
magnetic field strength in the Galactic halo goes down to reasonable
values of 2~$\mu$G.

\paragraph{External galaxies:} 
Using the energy equipartition condition the total magnetic field
strength in galaxies is $9\pm 3~\mu\rm G$ \citep{niklas_97a}. From the
linear polarization one can derive the regular magnetic field strength
as typically $1-5~\mu\rm G$ in the disk \citep{krause_09a}. In the
halo, the magnetic field is dominated by the regular component as
opposed to the disk. The regular component in the halo is of the order
of the disk component, if we neglect stronger fields that are found in
the interarm regions in the disk such as in NGC\,6946 \citep[][Beck,
this volume]{beck_07a}.

\section{Magnetic field structure in the Galaxy}
\label{s:milkyway}

\begin{figure}[h]
\includegraphics[width=\textwidth]{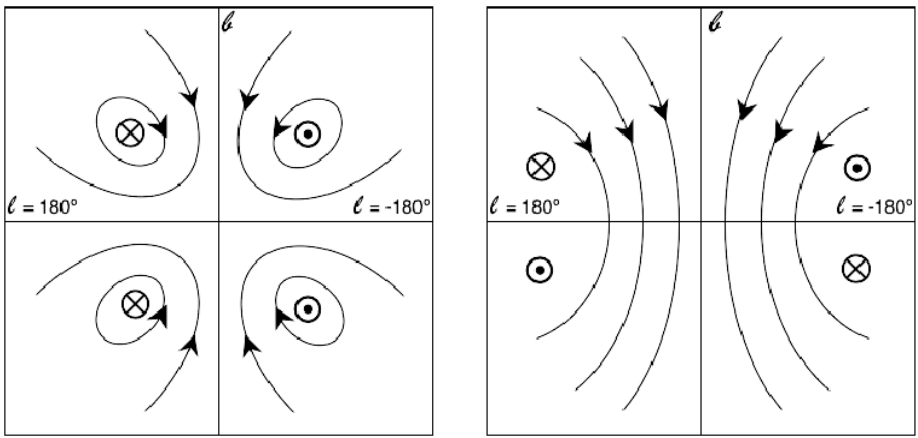}
\caption{Quadrupole (left) and dipole (right) large-scale magnetic
field configurations of the Milky Way as a function of Galactic
longitude and latitude. Magnetic field towards the viewer is denoted
by a dot, field away from the observer as a cross.}
\label{f:symm}
\end{figure}

Galactic magnetic fields are maintained and amplified by dynamo
action (see Brandenburg et al, this volume). The simplest model is that
of the $\alpha-\omega$ dynamo, which amplifies the radial magnetic
field component through differential rotation, and amplifies the
azimuthal and poloidal components of the field by turbulent loops
twisted by the Coriolis force. Although the $\alpha-\omega$ dynamo is
believed to act in the Sun \citep{o03}, it is considered unable to
sufficiently amplify galactic magnetic fields to observed values (eddy
diffusion time scales are much shorter than time scales for
amplification of the regular field, thereby suppressing turbulent
motions, and quenching the dynamo). Various solutions to this problem
are discussed in \citet{w02}.  For a flat disk-like galaxy which is
differentially rotating, mean-field dynamo theory predicts a
quadrupolar magnetic field configuration, where the direction of the
azimuthal magnetic field is the same above and below the plane, but
the direction of the vertical magnetic field component reverses with
respect to the plane. However, for a spherical, weakly rotating
structure -- such as a Galactic halo -- the dipolar configuration is
more easily excited, i.e.\ an azimuthal magnetic field with reversing
direction across the Galactic plane, while the vertical field is
directed in the same way above and below the plane (see
Figure~\ref{f:symm}).

\begin{figure}[h]
\begin{center}$
\begin{array}{ccc}
\includegraphics[width=0.3\textwidth]{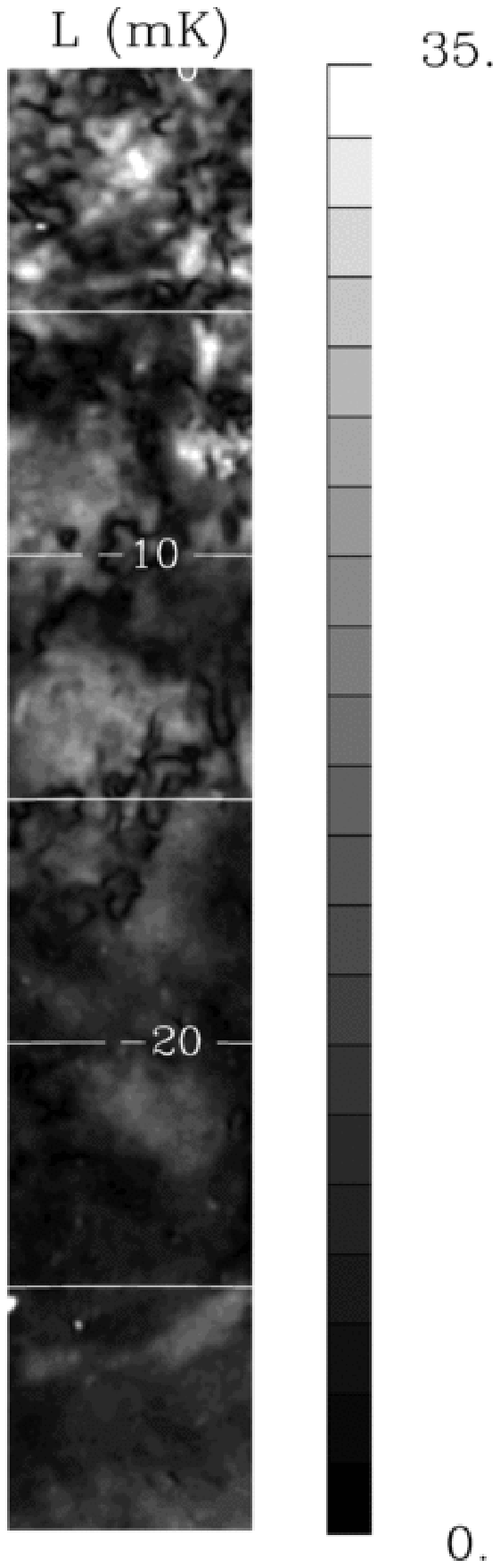} &
\includegraphics[width=0.3\textwidth]{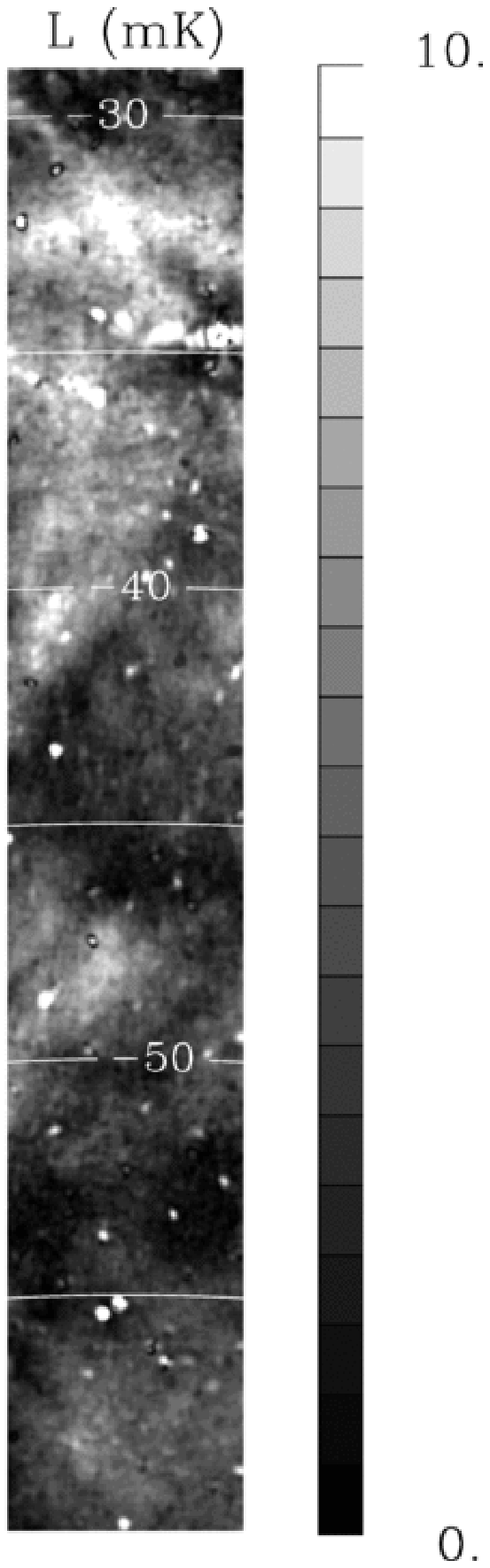} &
\includegraphics[width=0.3\textwidth]{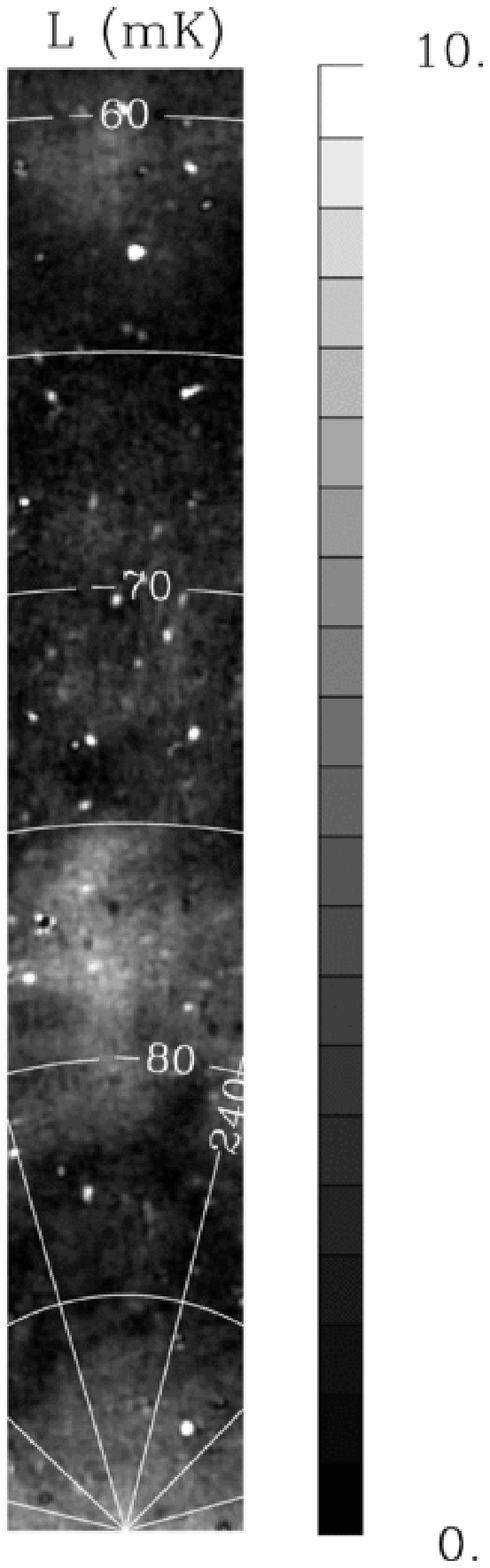}
\end{array}$
\end{center}
\caption{Linearly polarized intensity from the Parkes Galactic
Meridian Survey \citep{chm10} at 2.3~GHz shows small-scale structure
in linear polarization at all Galactic latitudes, indicating
small-scale structure in the electron-density weighted magnetic field.}
\label{f:pgms}
\end{figure}
\begin{figure}[tbhp]
\begin{minipage}[t]{0.5\textwidth}
   \resizebox{0.9\hsize}{!}{
   \includegraphics{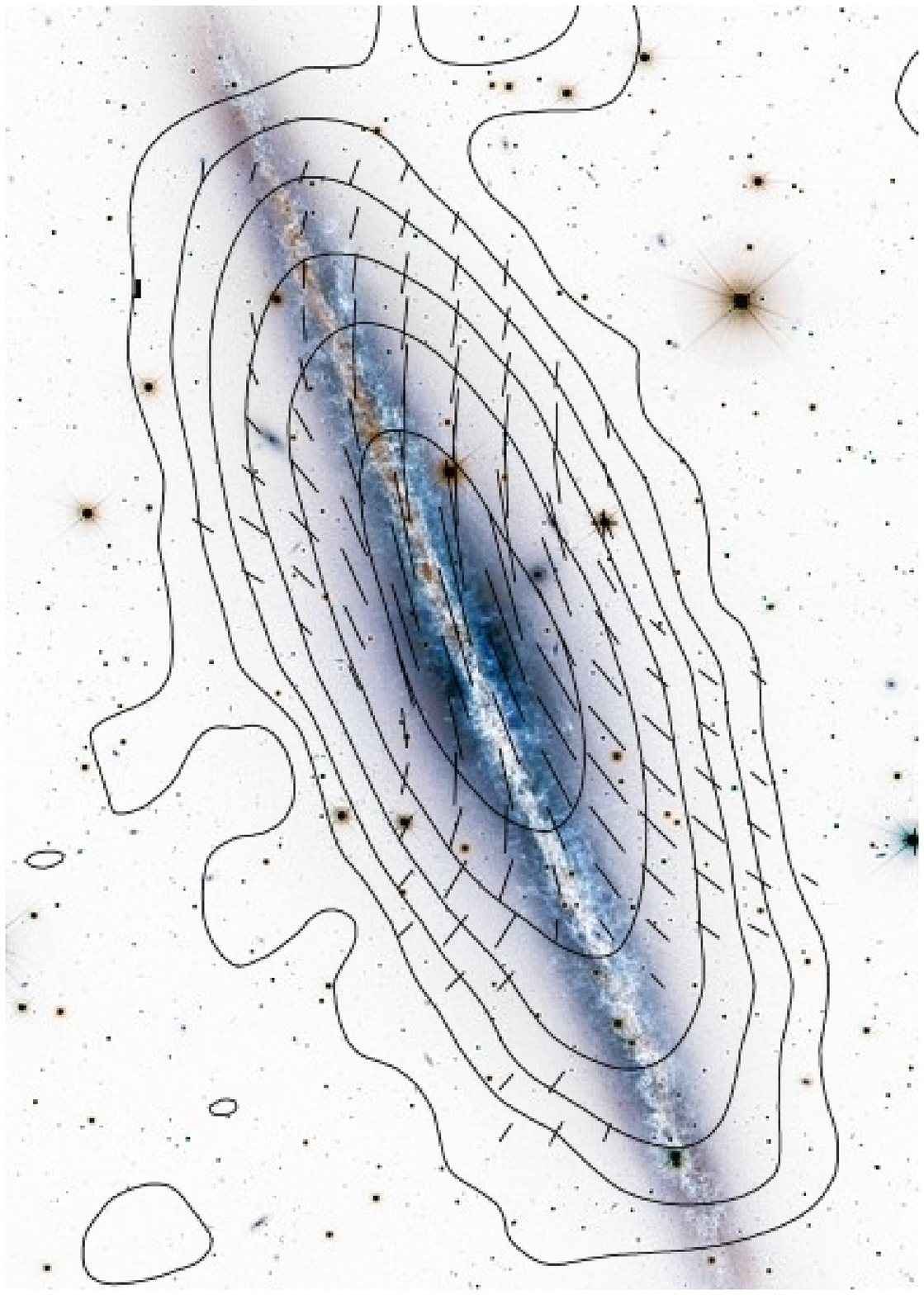}}
\end{minipage}
\hfill
\begin{minipage}[b]{0.49\textwidth}
 \caption{Radio continuum emission of the edge-on spiral galaxy NGC\,891
 at $\lambda$3.6~cm (8.35~GHz) with the 100~m Effelsberg telescope with
 a resolution of 84'' HPBW. The contours give the total intensity,
 the vectors the intrinsic magnetic field orientation (Copyright: MPIfR Bonn).
 The radio map is overlaid on an optical image of NGC\,891 from the
 Canada-France-Hawaii Telescope/(c)1999 CFHT/Coelum. From \citet{krause_09a}.}
 \label{fig:n891}
 \end{minipage}
\end{figure}

These two configurations can be observationally distinguished by the
directions of the azimuthal magnetic field component above and below
the Galactic plane, and by the directions of the vertical field
components above and below the plane. Firstly, measurements of Faraday
rotation of extragalactic sources show a pronounced anti-symmetric
``butterfly pattern'' with respect to the Galactic plane and the
meridian through the Galactic center \citep{sk80}, only visible in the
inner Galaxy (first and fourth Galactic quadrants). This has been
interpreted as a reversal of the direction of the azimuthal magnetic
field with respect to the Galactic plane indicative of a dipolar
magnetic field configuration and an A0 dynamo \citep{hmb97}, but can
possibly also be due to the influence of local structures on large
angular scales \citep{wfl10,sts11}. Secondly, a vertical component of
the Galactic magnetic field is predicted by \citet{bc90} to provide a
magnetic tension term that can balance other pressure components in
the Galactic disk and halo \citep[see also][]{am88}.  \citet{mao_10a}
tried to measure this vertical magnetic field component at high
latitudes. At the solar radius, they found a small vertical magnetic
field component of $B_z = 0.31 \pm 0.03~\mu$G towards negative
Galactic latitudes but not towards positive latitudes, indicating
smaller-scale structure in the halo magnetic field. These conclusions
agree with the structure in the X-ray halo \citep{f01} and with star
light polarization measurements.  \citet{tss09} include the North
Polar Spur region in their determination of vertical magnetic field
strengths and do find a small vertical magnetic field component $B_z =
-0.14 \pm 0.02~\mu$G, while their estimates towards the southern sky
agree with \citet{mao_10a}.

\citet{skh07} also find a small, varying vertical magnetic field
component at intermediate positive latitudes in the outer Galaxy.
\citet{dkh06} conclude from radio polarimetric measurements at high
Galactic latitudes that there must be significant variations in the
vertical component of the Galactic magnetic field on the level of
$1\mu$G, similar to the strength of the random magnetic field
component at high Galactic latitudes \citep{mao_10a}. However, the
studies by \citet{skh07} and \citet{dkh06} were performed at low
frequencies, which indicates that they may be probing mostly nearby
structure. The higher-frequency Parkes Galactic Meridian Survey at
2.3~GHz detects polarized emission from a larger path length and shows
that even towards the southern Galactic pole, there is still
small-scale structure in radio polarization, indicating structure in
the density-weighted Galactic magnetic field on scales of degrees
\citep[][see Fig.~\ref{f:pgms}]{chm10}.

Small-scale structure in the high latitude magnetic field is confirmed
by measurements of linear polarization of starlight. Starlight which
propagates through asymmetric dust grains aligned in a magnetic field
will become partially polarized with a polarization angle in the
average direction of the magnetic field perpendicular to the line of
sight in the dusty regions between the star and the observer.
Berdyugin and co-workers published a series of papers
\citep[e.g.][]{bt01,bth01,bpt04} on polarization directions of stars
at high latitude, which probe mostly the direction of the field
component parallel to the Galactic plane. They concluded (1) that this
magnetic field towards the southern Galactic pole is oriented towards
$l \approx 80^{\circ}$, in agreement with a spiral magnetic field, and
(2) that smaller-scale structure towards the northern Galactic pole
precluded any conclusions about the ordered field component there. As
the polarization percentages keep increasing towards higher distances
from the plane, out to their furthest stars at about 500~pc, they
concluded that Galactic magnetic fields reach at least to 500~pc
without significant decrease.

\begin{figure}[tbhp]
\begin{minipage}[t]{0.7\textwidth}
  \resizebox{0.9\hsize}{!}{\includegraphics{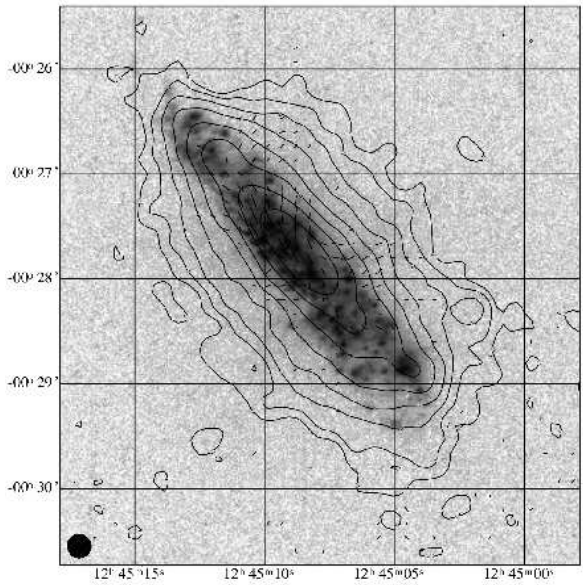}}
\end{minipage}
\begin{minipage}[b]{0.29\textwidth}
  \caption{Radio continuum emission of NGC\,4666 at $\lambda$6~cm from VLA
    observations with 14''
    HPBW. The vectors show the magnetic field orientation where the
    length is proportional to the polarized emission. The
    background shows H$\alpha$+N[II] emission, showing the spatial correlation of
    optical emission line and (polarized) radio synchrotron emission from the
    halo. From \citet{dahlem_97a}.}
 \label{fig:n4666}
\end{minipage}
\end{figure}
\begin{figure}[tbhp]
\begin{minipage}[t]{0.7\textwidth}
  \resizebox{0.9\hsize}{!}{\includegraphics{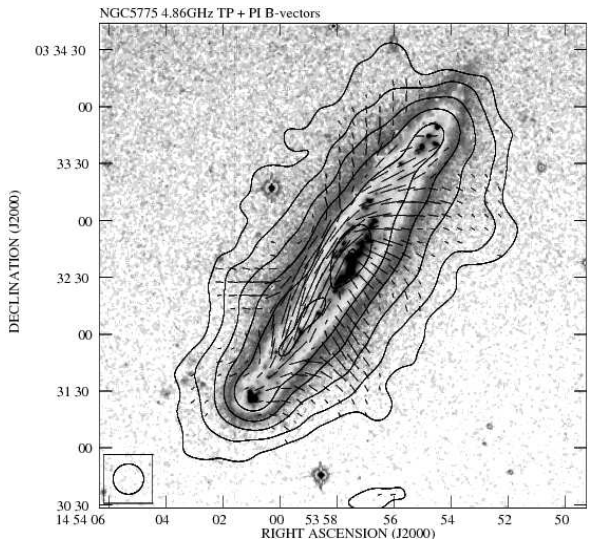}}
\end{minipage}
\begin{minipage}[b]{0.29\textwidth}
\caption{Radio continuum emission of NGC\,5775 at $\lambda$6~cm from VLA
  observations with 14'' HPBW. The vectors show the magnetic field orientation where the
    length is proportional to the polarized emission. The background shows
    H$\alpha$ emission. From \citet{tuellmann_00a}.}
 \label{fig:n5775}
\end{minipage}
\end{figure}
\section{Magnetic field structure in external galaxies}
\label{s:externals}
\subsection{Field orientation}
\begin{figure}[tbhp]
\resizebox{\hsize}{!}{\includegraphics{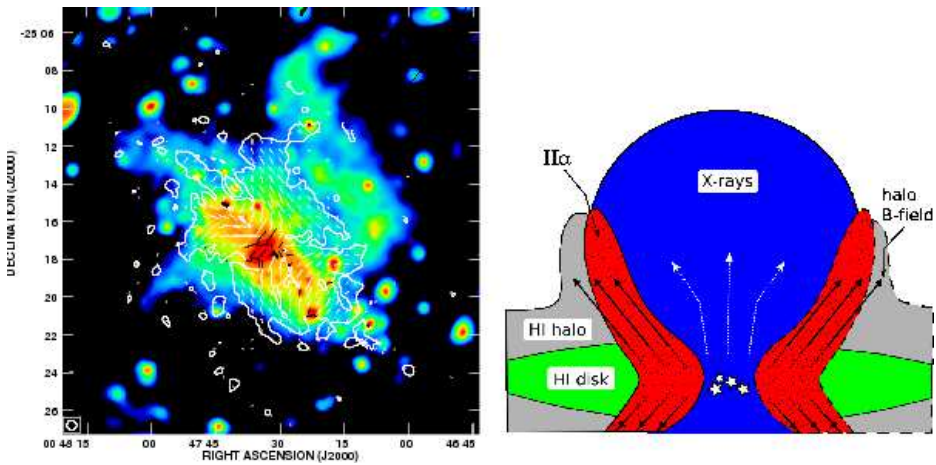}}
\vfill
\caption{Left: halo magnetic field in NGC\,253 overlaid on diffuse X-ray
  emission. Contours show the intensity of the polarized emission of the halo
  magnetic field. Vectors show the orientation of the halo magnetic field
  where the length is proportional to the polarized intensity. Right: proposed
  halo structure of NGC\,253. The superbubble, filled with soft X-ray
  emitting gas, expands into the surrounding medium (indicated by dotted lines
  with arrows). The halo magnetic field is aligned with the walls of the
  superbubble. Dashed lines denote components that are not (or only weakly)
  detected in the southwestern half of NGC\,253. Figures from
  \citet{heesen_09b}.}
\label{fig:n253_pol_XMM2}
\end{figure}

The magnetic field is mainly aligned parallel to the disk in regions close to
the mid plane. In the halo, the vertical component becomes stronger with
increasing height above the disk. Also the magnetic field is more ordered as
indicated by the higher polarization degree in the halo. Galaxies that have a
radio halo possess also vertical fields.
Several disk galaxies viewed edge-on show X-shaped magnetic fields in their
haloes as shown in Fig.~\ref{fig:n891} for the case of NGC\,891. The
orientation of the magnetic field aligns with the X in the four quadrants and
the polarized emission is enhanced there. Some galaxies have distinctive sites
with significant vertical magnetic field components that are referred to as
``radio spurs''. They correlate in position with filamentary H$\alpha$
emission sticking out from the disk into the halo. Examples for this are
NGC\,4666 \citep{dahlem_97a} and NGC\,5775 \citep{tuellmann_00a}. This is
illustrated in Figs.~\ref{fig:n4666} and \ref{fig:n5775}.
The reason for this structure is not yet clear. It could be related to the
flow of gas in the halo that has X-shaped structure as well
\citep{dallavecchia_08a}. The energy densities of the magnetic field and the
gas are comparable, so that interaction between the two is
possible. \citet{heesen_09b} showed that NGC\,253 also has an X-shaped
field.  This is in so far remarkable as the halo field can only be revealed after
subtraction of a model field for the disk,  because this galaxy has an inclination
angle of $78^\circ$ and is therefore not exactly edge-on. In the case of NGC\,253,
\citet{heesen_09b} suggested that the halo magnetic field may be enhanced by
compression due to expanding hot gas in the halo that is visible as a conical
outflow in X-ray emission (see Fig.\,\ref{fig:n253_pol_XMM2}). Consequently,
radio spurs should align with the boundary of the outflow cone and the
magnetic field be roughly tangential to it.

\begin{figure}[tbhp]
\begin{minipage}[t]{0.7\textwidth}
   \resizebox{0.9\hsize}{!}{
   \includegraphics{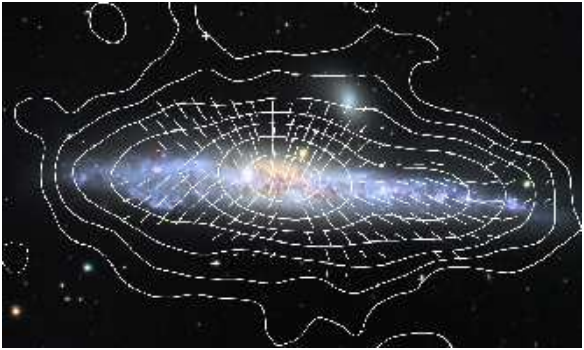}}
\end{minipage}
\hfill
\begin{minipage}[b]{0.29\textwidth}
  \caption{Radio continuum emission of NGC\,4631 at $\lambda$3.6~cm with 84''
    HPBW from observations with the 100~m Effelsberg telescope. The vectors
    show the magnetic field orientation where the length is proportional to
    the polarized emission. The background is an optical image taken at the
    Misti Mountain Observatory (Copyright: MPIfR Bonn). From \citet{krause_09a}.}
 \label{fig:n4631}
 \end{minipage}
\end{figure}
An example of non-X-shaped fields is NGC\,4631 with a rather more dipolar field
structure, where the vertical component of the magnetic field dominates as
shown in Fig.\,\ref{fig:n4631}. This may be a case where the field is
dominated by an interaction with a nearby companion or where differential
rotation is weak and thus the A0 dynamo mode is excited
\citep{golla_94a}.  Deviations from the X-shaped field structure are also
found for galaxies which do not have star formation driven outflows. NGC\,4258
\citep{krause_04a} and NGC\,4569 \citep{chyzy_06a} are examples for jet
magnetic fields. There the magnetic field is vertically aligned to the disk and
parallel to the outflow in the jet.
Vertical magnetic fields are also indirectly detected. M\,82 has vertical
filaments and gaps in the radio continuum emission at $\lambda$20\,cm with a
width of 100\,pc \citep{reuter_92a,wills_99a}. Vertical fields were detected
with radio continuum polarimetry in this galaxy albeit at a lower resolution,
which likely prevented the detection of filaments
\citep{reuter_94a}. \citet{wills_99a} tried to connect the filaments to
chimney structures, i.e.\ blow outs by clustered supernovae, but the outflow
visible in CO line emission was only related to a weaker minimum in the
continuum emission. \citet{duric_98a} found vertical ``tentacles'' of a flat
spectral index component protruding from the disk into the halo in NGC\,5775. If they
are interpreted as outflows of young cosmic-ray electrons, there needs to be
some vertical alignments for which magnetic fields are the favored
mechanism. There are other indicators of vertical fields like vertical dust
filaments in NGC\,253 \citep{sofue_94a}. These filaments are thin ($\le 50~\rm
pc$) compared to the resolution of the available radio maps and are thus far not
detected in the continuum or polarized emission.
Nuclear outflows are a good site to study filaments in the radio continuum,
because they are very bright and allow us to use high resolution. Although
they are not located in the halo, they are at the base of the conical outflows
discussed above. They may be also a source for vertical fields. An example for
nuclear outflow is NGC\,3079 which has a prominent bowl-shaped structure
visible as H$\alpha$ filaments. The polarized emission is concentrated in
filaments bordering the outflow cone and is tangential to it
\citep{cecil_01a}. Similarly, the outflow cone in NGC\,253 is bordered by
continuum filaments with a scale height of $150\pm 20~\rm pc$ with the field
orientation parallel to the filaments \citep{heesen_11a}. The magnetic field
may collimate the nuclear outflows and thus explain the small opening angles
$\sim 30^ \circ$ of the cones.
Whereas interaction in the interstellar medium can explain the distribution of
the fields in filaments, it does not \emph{generate} vertical magnetic
fields. For this the galactic dynamo is a favorite process. Vertical magnetic
fields may also be generated by the Parker instability where cosmic-ray gas
inflates magnetic field lines that reconnect and give rise to the fast Parker
dynamo \citep{parker_92a}. These magnetic fields would be expected to be
anisotropic with frequent field reversals. This configuration is expected to
have reversals in the line-of-sight component which can be detected in Faraday
rotation.
\subsection{Field direction} 
The halo rotation measures (RMs) are
difficult to measure, because they can overlap with the RMs in
the disk which dominate the polarized emission. Recent papers have
tried to separate the contributions from the disk and halo by using
models of the magnetic field structure. \citet{heesen_09b} found in
NGC\,253 a magnetic field that points outward on the near side of
the halo, but the data is not conclusive for the halo behind the
bright polarized emission in the disk. The field direction has some
consequence for the dynamo models because in the presence of a galactic
wind the parity of the disk and halo field are expected to be
identical \citep{moss_10a}.
\citet{braun_10a} explained the asymmetry of the polarized emission with
respect to the receding and approaching side of the galaxy with halo
fields. In their sample of 21 galaxies detected in polarization
\citep{heald_08a} they found directions that resemble both the quadrupolar
type of fields and dipolar radially directed fields. The magnetic field
structure in the halo is also of interest for angular momentum
transport. If the field lines are stiff and we do not measure any azimuthal
component then we know that we are below the Alfv\'{e}nic point. The higher up
the Alfv\'{e}nic point the more angular momentum is transported from the disk
into the halo \citep{zirakashvili_96a}.
The Parker fast dynamo would see RM reversals as part of Parker loops. Such a
reversal was observed in the nuclear outflow in NGC\,3079 \citep{cecil_01a}. In
NGC\,1569 the vertical magnetic field borders an outflow bubble and is also
accompanied with a field reversal \citep{kepley_10a}. In principle it is
possible to distinguish between the anisotropic and the isotropic component of
the halo field by comparing the field strengths from the RM and 
equipartition. This was attempted by \citet{heesen_09b} where they found the
field component from the RM to be only 50\% lower than that from
equipartition. This would mean that the halo field has only a small
anisotropic component generated by Parker loops.
\section{Interaction of the magnetic field with the interstellar
medium}
\label{s:interaction}

Interstellar magnetic fields are in constant interaction with the other
components in the interstellar medium. Magnetic fields are mostly directly
frozen into the ionized gas component, but also closely interact with the
neutral gas through charge exchange \citep[for an example, see][]{mdg06}. On
large scales, the magnetic field energy density is comparable to the cosmic
ray energy density and the turbulent gas energy density (see Heiles \&
Haverkorn this volume; and Beck, this volume). This means that any of these
components can only be fully understood if the interaction with the other
components is taken into account.
Therefore, gas dynamics is greatly influenced by magnetic
fields. Numerical simulations show that expansion of supernova
remnants in a magnetized medium will be anisotropic \citep{swo09} and
slower \citep{fmz91} than in the non-magnetized
case. Magneto-hydrodynamic turbulence can be a significant source of
heating in the ISM \citep[e.g.][]{se04}, as can magnetic reconnection
\citep{zlb97}.

Magnetic fields play a large role in the hydrostatic equilibrium of
the Milky Way. Thick gas disks, as observed in the Milky Way, with
purely azimuthal magnetic fields, have a tendency to become unstable
\citep[e.g.][]{p69}. \citet{bc90} show that magnetic tension in a
small vertical component of the magnetic field may stabilize this
configuration and so help to keep the thick gas layer from collapsing
onto the disk.

Cosmic rays are tightly coupled to galactic magnetic fields. These
charged particles gyrate around magnetic field lines with a Larmor
radius depending on their energy and the magnetic field strength.
Therefore, magnetic fields effectively trap all except the most
energetic cosmic rays in galaxies and fully determine their
trajectories. Also, magnetic fields are a major factor in
acceleration of these charged particles to their relativistic
velocities. For more details, see Aharonian et al. (this volume).

\begin{acknowledgements}
The authors want to thank the International Space Science Institute
for support and organisation of an excellent meeting. We are also
indebted to Rainer Beck and Elias Brinks for fruitful discussions and
useful comments which improved the paper.
\end{acknowledgements}

\end{document}